\documentclass[11pt,a4paper]{article}
\pdfoutput=1
\usepackage{jheppub}
\usepackage{latexsym,amsmath,amssymb}
\usepackage{graphicx,subcaption}
\usepackage[table,usenames,dvipsnames]{xcolor}
\usepackage{hyperref}
\definecolor{azure(colorwheel)}{rgb}{0.0, 0.5, 1.0}
\hypersetup{colorlinks=true,linkcolor=blue,citecolor=azure(colorwheel),pdfborder={0 0 0}}
\usepackage[utf8]{inputenc}
\usepackage{feynmp}
\newcolumntype{P}[1]{>{\centering\arraybackslash}p{#1}}

\makeatletter
\newcommand*{\rom}[1]{\expandafter\@slowromancap\romannumeral #1@}
\makeatother

\newcommand{\ignore}[1]{}

\begin{document}
\title{
  Inching toward the QCD Axions with Axion Magnetic Resonance in Helioscopes
}

\author[a]{Hyeonseok Seong,}
\author[bc]{Chen Sun}
\author[d]{and Seokhoon Yun}

\affiliation[a]{Deutsches Elektronen-Synchrotron DESY, Notkestr. 85, 22607 Hamburg, Germany}
\affiliation[b]{Theoretical Division, Los Alamos National Laboratory, Los Alamos, NM 87545, USA}
\affiliation[c]{Abdus Salam International Centre for Theoretical Physics, Strada Costiera 11, 34151, Trieste, Italy} 
\affiliation[d]{Particle Theory  and Cosmology Group, Center for Theoretical Physics of the Universe, Institute for Basic Science (IBS), Daejeon, 34126, Korea}

\emailAdd{hyeonseok.seong@desy.de}
\emailAdd{csun@ictp.it}
\emailAdd{seokhoon.yun@ibs.re.kr}

\abstract{
Utilizing a helical magnet profile to enhance axion-photon conversion showed great promise in laboratory searches for high axion masses.
We extend the mechanism, known as the axion-magnetic resonance (AMR), from laser experiments to axion helioscopes and demonstrate its potential in covering QCD axion parameter space.
Specifically, we apply AMR to the CAST experiment legacy, make projections for the upcoming IAXO experiment, and assess its implications for both axion-like particles and QCD axions.
We observe considerable improvement in the experiment's sensitivity reach in all cases.
}

\date{\today}


\makeatletter
\gdef\@fpheader{}
\makeatother

 \preprint{
DESY-24-124, LA-UR-24-28820, CTPU-PTC-24-25}
 \preprint{
 \begin{minipage}[t]{\textwidth}
\raggedleft
DESY-24-124 \\
LA-UR-24-28820\\
CTPU-PTC-24-25
\end{minipage}
}

\maketitle


\section{Introduction}
\label{sec:introduction}

\subsection{Overview}

The strong CP problem is notable among the fine-tuning problems in the standard model, which include the gauge hierarchy problem, the cosmological constant problem, and the strong CP problem itself.
This problem not only involves the quantum chromodynamics (QCD) sector but also connects with the flavor sector through the parameter $\bar{\theta} \equiv \theta_{\rm QCD} + {\rm arg}\,{\rm det}M_u M_d$, where $\theta_{\rm QCD}$ is the bare CP-violating parameter in the QCD Lagrangian, and $M_{u,\,d}$ are the quark mass matrices.
Furthermore, the strong CP problem is closely related to low-energy dynamics concerning the QCD axion solution.
One approach to addressing this problem is to promote the $\bar{\theta}$ parameter to a dynamical field that transforms under a global $U(1)$ symmetry \cite{Peccei:1977hh,Weinberg:1977ma,Wilczek:1977pj}, so-called the QCD axion.
Since the CP-conserving vacuum, where $\bar{\theta} = 0$, has a minimum in energy \cite{Vafa:1984xg}, the QCD axion relaxes to this CP-conserving vacuum over the cosmological history.
In addition to the conciseness of its solution, the QCD axion has numerous phenomenological consequences, which are comprehensively reviewed in \cite{Kim:2008hd,Marsh:2015xka,Irastorza:2018dyq,DiLuzio:2020wdo,Choi:2020rgn,Caputo:2024oqc,Ringwald:2024uds}.
With the recent surge of interest in axion-like particles (ALPs) expected in UV theories such as extra-dimensions or string theory \cite{Witten:1984dg,Svrcek:2006yi,Demirtas:2021gsq,Gendler:2023kjt,Choi:2024ome,Gendler:2024adn}, finding the QCD axion remains one of the most crucial tasks in ongoing phenomenological studies.

The importance of laboratory experiments that are independent of the dark matter abundance in the search for the QCD axion cannot be overstated.
Although astrophysical and cosmological studies \cite{Marsh:2015xka,Caputo:2024oqc} put efforts towards the QCD axion parameters, and axion haloscopes \cite{ADMX:2021nhd,CAPP:2024dtx,HAYSTAC:2023cam,Alesini:2020vny,Quiskamp:2022pks} even reach the QCD axion lines represented by the KSVZ \cite{Kim:1979if,Shifman:1979if} and DFSZ \cite{Dine:1981rt,Zhitnitsky:1980tq} axion models for some mass windows around $m_a\sim \mu {\rm eV}$, those laboratory experiments not relying on the axion dark matter would provide the most definitive evidence for the QCD axion with minimal uncertainties.
The QCD axions of the KSVZ and DFSZ models remain largely unexplored over the range $m_a\simeq 10^{-5}\,{\rm eV} - 10^0\,{\rm eV}$.
Given that, we have proposed a resonance from a magnetic field configuration with a specific length scale \cite{Seong:2023ran}, named axion magnetic resonance (AMR).
The length scale of a magnetic field configuration, e.g., the period of spatially-varying magnetic fields, acts as an effective photon mass that cancels the axion mass and leads to the resonant conversion between axions and photons.
This resonance allows light-shining-through-the-wall (LSTW) experiments, such as ALPS II \cite{Bahre:2013ywa,Ortiz:2020tgs}, to extend their axion mass coverage closer to the QCD axion lines.\footnote{Branching from the original conception of the LSTW experiment \cite{VanBibber:1987rq,Anselm:1985obz,Anselm:1985oaj}, numerous variations have been proposed, including the use of an alternating magnet design \cite{VanBibber:1987rq,Arias:2010bha,Arias:2010bh}, optical resonators that inspired the ALPS II experiment \cite{Hoogeveen:1990vq,Sikivie:2007qm,Mueller:2009wt}, phase-shift plates \cite{Jaeckel:2007gk}, superconducting radio-frequency cavities \cite{Janish:2019dpr}, a collider setup \cite{Kling:2022ehv}, and the ultimate configuration of the LSTW experiment \cite{Hoof:2024gfk}.
In a slightly different but closely related context, ideas such as measuring a phase shift with varying magnetic field profiles \cite{Zarei:2019sva,Sharifian:2021vsg} and ALP production via synchrotron radiation \cite{Yin:2024rjb} have also been explored.
}

In this work, we propose to apply the newly identified resonance, AMR, to the axion helioscopes, e.g., CAST, BabyIAXO, IAXO, and IAXO+ \cite{Kotthaus:2005zg,Kuster:2007ue,CAST:2017uph,IAXO:2019mpb,IAXO:2020wwp}, by leveraging a helical magnetic field profile.\footnote{A sinusoidal magnetic field profile for axion helioscopes was discussed in Sikivie's pioneering paper \cite{Sikivie:1983ip}. We reformulate this idea by focusing on a helical magnetic field profile, which maximizes axion-photon conversion at the specific axion mass where AMR occurs.}
We aim to overcome the axion mass suppression caused by the growing mismatch of the dispersion relations of axions and photons, which starts to appear around $m_a\simeq 10^{-2}\,{\rm eV}$, and to extend the coverage towards the various realizations of the QCD axion lines, for $m_a\simeq \mathcal{O}(10^{-1})\,{\rm eV}$.
It is worth noting that the AMR in axion helioscopes manifests differently compared to that in LSTW experiments, in the context of the energy spectrum.
The LSTW experiments use a mono-chromatic energy spectrum, but the axion helioscopes use a continuum energy spectrum of solar axions.
Moreover, in the DFSZ axion model, the axion-electron coupling contribution to solar axions dominates over the axion-photon coupling contribution.
We thus present our constraints for the QCD axion models by taking this axion-electron coupling contribution into account.
In addition, other channels through the axion couplings to electrons and nucleons are thoroughly examined as well.


\subsection{Lightning Review of AMR}
\label{sec:AMRreview}

Many axion models predict an interaction between axions and photons through the following anomalous Lagrangian term:
\begin{align}
    \mathcal L
    & \supset
    \frac{g_{a\gamma}}{4} a F_{\mu\nu} \tilde F^{\mu\nu},
\end{align}
where $\tilde F^{\mu\nu} = \epsilon^{\mu\nu\alpha\beta} F_{\alpha\beta}/2$ is the dual of the field strength tensor $F_{\alpha\beta} = \partial_\alpha A_\beta - \partial_\beta A_\alpha$.
For relativistic axions and photons, the equation of motion reads
\begin{align}
     i \partial_z
    \begin{bmatrix}
      \gamma_x \\
      \gamma_y\\
      a
    \end{bmatrix}
  & =
    \frac{1}{2\omega}
    \begin{bmatrix}
      0 & 0 & g_{a\gamma}  \omega B_x(z) \\
      0 & 0 & g_{a\gamma} \omega B_y(z)\\
      g_{a\gamma} \omega B_x(z) & g_{a\gamma}  \omega B_y(z)  & m^2_{a}
    \end{bmatrix}
    \begin{bmatrix}
      \gamma_x \\
      \gamma_y\\
      a
    \end{bmatrix},
    \notag
\end{align}
where the $z$-axis is the propagating direction, and $\gamma_{i=x,y}$ and $a$ denote the photon and axion states, respectively. $\omega$ is the energy of the axion and the photon, and $m_a$ is the axion mass.
$B_{i=x,y}(z)$ represents the $i=x,y$ components of the background magnetic field, which vary along the $z$-direction.
This set of equations describes the mixing between axions and photons in the background magnetic fields.
As two-level quantum systems experience the Rabi oscillation, this axion-photon system can experience a resonantly enhanced conversion to one another if the mixing term has an explicit dependence on time.\footnote{Since both particles are relativistic, time and distance are not distinguished here.} Ways to induce this resonance include utilizing an external magnetic field with a harmonic spatial profile in its magnitude, a spatial helical profile in its orientation~\cite{Sikivie:1983ip,Seong:2023ran}; or a temporal oscillation~\cite{Sharifian:2021vsg,Seong:2023ran}.
Let us take the spatial helical profile as an example and parametrize the magnetic field components $B_x=B\sin\dot\theta z$, $B_y=B\cos\dot\theta z$ with a constant rotation frequency $\dot\theta$ and a magnitude $B$.
If $m_a^2>g_{a\gamma}\omega B$, which is indeed the case of our interest, the axion-to-photon conversion probability is given by
\begin{align}
\label{eq:conversion-prob-varyingB}
P_{a{\rightarrow} \gamma}
& \simeq 
\sum_{i=\pm}\frac{(g_{a\gamma}B/\sqrt{2})^2}{\Delta_i^2}
  \sin^2 \left (\frac{\Delta_i z}{2} \right) \,
\end{align}
with
\begin{align}
\label{eq:Delta-p-m}
\Delta_\pm = \sqrt{\left( m_{a}^2/2\omega \pm \dot{\theta}\right)^2 + \left(g_{a\gamma} B/\sqrt{2}\right)^2} \, ,
\end{align}
where we sum over the two photon helicities. In the heavy axion mass regime where the axion-photon oscillation length $\sim 4\pi \omega/m_a^2$ becomes shorter than the size of the magnetic field domain $L_{\rm B}$, the axion-photon conversion is suppressed by $m_a^{-4}$, and we observe that, at the specific energy of $\omega_a \equiv |m_a^2/(2\dot \theta)|$, the conversion probability is significantly enhanced due to the cancellation between $m_a^2/(2\omega_a)$ and $\pm\dot\theta$.

We can compute the width of the resonance in terms of the photon energy. First, let us compare the case of a perfect resonance, $\omega = \omega_a$ with an imperfect resonance $\omega = \omega_a + \Delta \omega$. 
We require the latter to have a conversion probability of half of the resonance, hence, $\Delta \omega$ can be interpreted as the half resonance width. More precisely: 
\begin{align}
\frac{P_{a\rightarrow\gamma}(\omega_a)}{P_{a\rightarrow\gamma}(\omega_a+\Delta \omega)} = 2.
\end{align}
We get $\Delta \omega \approx (2 /\dot\theta L_B) \omega_a $.
More practically, we can require the enhancement is actually an enhancement. Imagine that we choose a helical frequency $\dot\theta$ that poorly matches $ m_{a}^2/2\omega$. The presence of $\dot \theta$ itself can impede the axion-photon conversion. By requiring the conversion probability is not worsened by the presence of a $\dot \theta$, we get the width during which the experimental reach is enhanced by introducing a variation of the magnetic field.
\begin{align}
    P_{a\rightarrow \gamma} (\omega; \dot\theta) > 
    P_{a\rightarrow \gamma} (\omega; 0)
\end{align}
This simply gives us $\delta \omega \approx \omega_a$. This is usually much wider than $\Delta \omega$ since the resonance usually enhances $P_{a\rightarrow \gamma}$ by a few orders of magnitude. 
This can also be translated to the range of axion mass an AMR setup is sensitive to. In laser-based setups (e.g., LSTW), where a monochromatic frequency around the optical wavelength range $(\sim {\rm eV})$ is considered, AMR impacts only the specific axion mass $m_a = \sqrt{2\omega|\dot{\theta}|}$.~\footnote{In Ref.~\cite{Seong:2023ran}, a possible scanning method is proposed by exploiting a discrete domain methodology, which improves the original proposal based on alternating wigglers~\cite{Sikivie:1983ip}.}

Unlike the mono-energetic axion flux produced from high intensity lasers, solar axions have a continuum energy spectrum. This allows a single helical setup, \textit{i.e.} fixed $\dot \theta$, to cover a wider axion mass range as long where the resonance condition $m_a^2/(2\omega)=|\dot\theta|$ is satisfied at different energies for axions with different masses.
We argue in this work that AMR can be equally efficient with continuum axion flux such as the solar axions. In particular, a fixed $\dot\theta$ will be able to enhance the sensitivity of a helioscope for a range of axion masses.

\section{Axion Magnetic Resonance at Helioscopes}
\label{sec:projections}

In this section, we discuss the effect of AMR on helioscopes when only the axion-photon coupling $g_{a\gamma}$ is present, which is the minimal standard setup.
Interestingly, AMR induces a distinguishable spectral shape for the axion-photon conversion probability compared to the conventional constant background magnetic field setup.
We will examine how AMR improves the sensitivities of helioscopes.

\subsection{Axion Magnetic Resonance with Continuum Flux}
\label{sec:reso-with-extend}


A key characteristic feature of solar axions is their continuous flux spectrum in the higher frequency range of $\mathcal{O}({\rm keV})$, corresponding to the core temperature of the Sun.
For simplicity, we first take the solar axion flux only from the Primakoff process as the most conservative approach. We will impose more theory consistency requirements in QCD models and include the model-dependent contributions to the flux in Sec.~\ref{sec:qcd-axion}.
The spectral fitting expression for the solar axion spectral flux is given by~\cite{CAST:2007jps,CAST:2017uph}\footnote{The solar model dependence of solar axions is well-established \cite{CAST:2007jps}.}
\begin{align}
\Phi_{a, P}
& = 
6.02 \times 10^{10}
\left (\frac{g_{a\gamma}}{10^{-10}  
\,\mathrm{GeV}^{-1}}
\right) ^2
\frac{\tilde \omega^{2.481}}{e^{\tilde \omega/1.205}} 
~~
[\mathrm{cm}^{-2}\, \mathrm{s}^{-1} \,
\mathrm{keV}^{-1}],
\label{eq:primakoff}
\end{align}
where $\tilde \omega = \omega/(1\,{\rm keV})$ is the photon energy in units of keV.
\begin{figure}[t]
  \centering
  \includegraphics[width=0.6\linewidth]{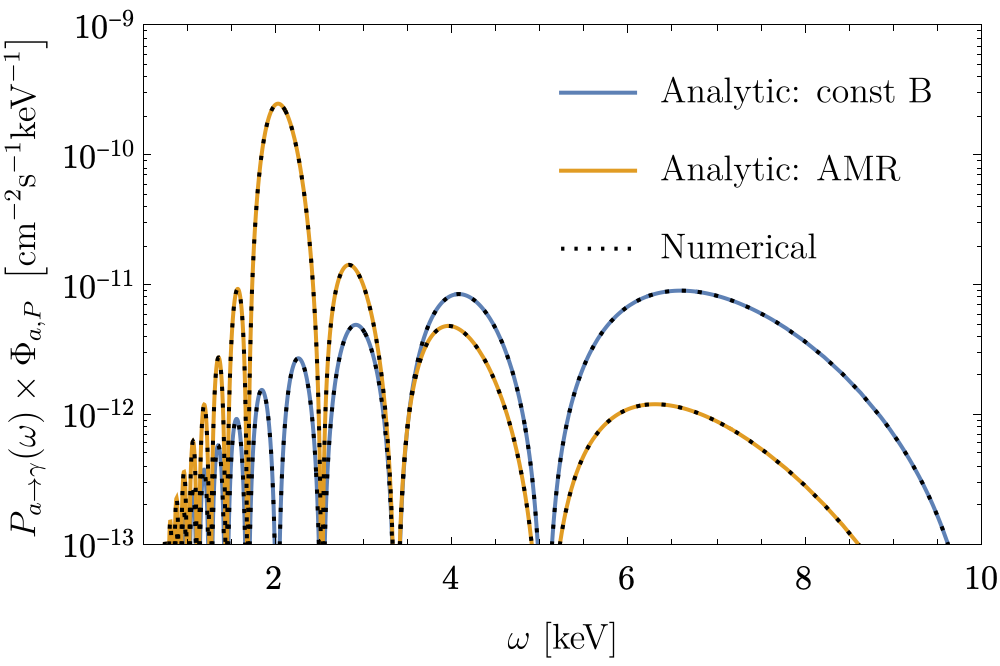}
  \caption{Photon signal converted from solar axions. In the constant $B$ case (blue), we assume $B=2\,\mathrm{T}$ spanning $10\,\mathrm{m}$. We fix $m_a = 0.05\,\mathrm{eV}$ with coupling $g_{a\gamma} = 3 \times 10^{-11}\,\mathrm{GeV}^{-1}$. In the case of AMR (orange), all are the same except that $B$ field has a helical profile with a period of $2$ meters, i.e., $\dot{\theta} = 2\pi/(2\,{\rm m})$. We verify the semi-analytic solution (solid curves, see Eq.~\eqref{eq:conversion-prob-varyingB}) with the full numerical solution (dotted) and found good consistency.
  }
  \label{fig:benchmark-flux}
\end{figure}
We highlight three advantages in applying AMR to solar axion experiments:
\begin{enumerate}
    \item 
Spanning an order of magnitude, the broad energy range relaxes the condition for the axion mass to satisfy AMR. Even in a fixed $\dot{\theta}$ setup, the improvement of experimental sensitivities through AMR happens at different energies for different axion masses where $m_a^2/2\omega = \dot \theta$ is satisfied.
\item The AMR-induced signal is more distinguishable from the background, which comprises rather {flat} dark counts. Axion helioscope experiments focus on distinguishing the axion-induced X-ray signal from these {smooth} dark count backgrounds~\cite{CAST:2017uph}. The AMR induced signal has stronger energy dependence allowing the experiments to better distinguish them from the flat background. 
\item AMR-induced signal allows a better suppression of the background should detectors with high energy resolution be adopted in the future.
This is because that, in the extreme scenario with large $m_a$ and large $\dot\theta$, signal X-ray lies in a small frequency band. Detectors with high resolution could enjoy a better energy cut that reduces the background without hurting the AMR-induced signals. 
\end{enumerate}

In Fig.~\ref{fig:benchmark-flux}, we show verify these three points by comparing the axion-induced X-ray signal with the signal resulted from a conventional homogeneous magnetic field. We see that the AMR-induced signal peaks around the AMR energy $\omega \sim \omega_a = m_a^2/(2\dot \theta)$. 

\subsection{Impact of AMR on Helioscope Sensitivity}
\label{sec:amr-at-helioscopes}
\begin{figure}[h]
  \centering
  \includegraphics[width=.6\linewidth]{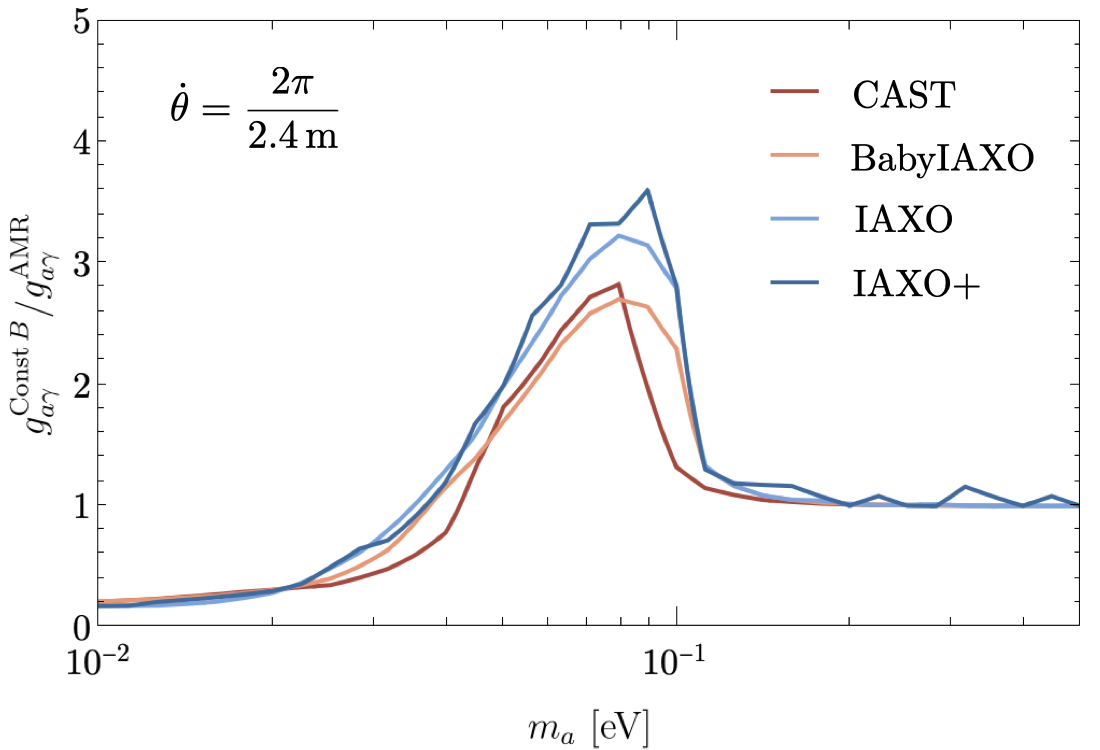}
  \caption{We show the amount of enhancement in four experimental setups due to twisting the magnetic profile, including CAST, BabyIAXO, IAXO, and IAXO+. The helical twist has a period of $2.4$ meters and we maintain the same specifications including the baseline length and magnetic strength to be the same as the current design of each experiment. 
  }
  \label{fig:xi-ratio-bound}
\end{figure}



Because AMR both enhances the signal flux and generates a unique spectral shape of the signal flux depending on the axion mass, it leads to improved sensitivities at the helioscope experiments. 
We define the enhancement factor, $\xi(m_a)$, to be
\begin{align}
    \label{eq:enhancement-factor}
    \xi(m_a)
    & \equiv
    \frac{g_{a\gamma}^{{\rm Const}\,B}(m_a)}{g_{a\gamma}^{\rm AMR}(m_a)},
\end{align}
where the denominator (numerator) is the experimental bounds on $g_{a\gamma}$ with (without) AMR, respectively. Next, we describe the analysis we performed for CAST, BabyIAXO, IAXO, and IAXO+.

We perform a simplified analysis {based on the extended maximum likelihood method} to derive the experimental bounds throughout this paper (see App.~\ref{sec:fit} for details of the extended maximum likelihood method).
We use the same log-likelihood function as in \cite{CAST:2017uph}:
\begin{align}
    {\rm ln}\,\mathcal{L}
    =&
    \sum_i 
    N^i_{\rm exp} \ln \, N_{\rm th}^i
     - N^i_{\rm th},
\end{align}
where the theory prediction ($N_{\rm th}$), the experimental observation ($N_{\rm exp}$), and the averaged signal flux are given in bin $i$ as follows.
\begin{align}
    N_{\rm th}^i
    & = N_{\rm sig}^i + N_{\rm bkg}^i 
    = \Phi_{\rm sig}^i (A  \epsilon_o \epsilon_d)\, \Delta \omega (\epsilon_t T)
    + \Phi_{\rm bkg}^i a \, \Delta \omega 
    \,(\epsilon_t T),
    \\
    N_{\rm exp}^i & = N_{\rm bkg}^i 
 = \Phi^i_{\rm bkg}\, a \, \Delta \omega\, \epsilon_t T,
 \label{eq:Nexpassump}
    \\
    \Phi_{\rm sig}^i 
    & = \frac{1}{\Delta \omega}
    \int_{i-{\rm th}\,{\rm bin}}
    \,
    d\omega \,
    P_{a\rightarrow \gamma}(\omega)
    \Phi_a
    .
\end{align}
Here we use the constant dark counting flux $\Phi_{\rm bkg}$ as given in Table.~\ref{tab:helioscope-specs}.
Here, we assume the observed flux $\Phi_{\rm exp}^i$ is equal to the dark count flux $\Phi_{\rm bkg}^i$, which implies null signals at detection.
To calculate the posterior probability distribution, we use a flat prior for positive values of $g_{a\gamma}$.\footnote{{As a sanity check, we also allow $g_{a\gamma}^4$ to be negative in our fit and confirm that the best-fit value is consistent with $g_{a\gamma}^4=0$ around the 1-$\sigma$ confidence level.}}
Since the $g_{a\gamma} = 0$ is consistent with the backgrounds-only assumption and maximizes the likelihood function, we can derive constraints on $g_{a\gamma}$ at the 95\% confidence level by integrating the posterior probability distribution from $g_{a\gamma}=0$ to $g_{a\gamma}=g_{a\gamma}^{95\%}$ at which the enclosed area under the probability distribution function becomes 95\% of the total area. 
We repeat this procedure for each experiment, including CAST, BabyIAXO, IAXO, and IAXO+, with specifications provided in Table~\ref{tab:helioscope-specs}.

\begin{table}[]
    \centering
    \begin{tabular}{l c c c c c}
    \hline
         Quantity & CAST & BabyIAXO & IAXO & IAXO+&  Unit\\
    \hline
         energy resolution $\Delta \omega$ & 1 & 1 & 1 & 1 & keV \\
         lower energy threshold $\omega_{\rm min}$ & 2 & 0.1 & 0.1 & 0.1 & keV\\
         upper energy threshold $\omega_{\rm max}$ & 7 & 10 & 10 & 10 & keV\\
         magnetic flux density $B$ & 9 & 2 & 2.5 & 3.5 & Tesla\\
         baseline length $L$ & 9.26 & 10 & 20& 22& meter\\
         detector efficiency $\epsilon_d$ & 0.6 & 0.7 & 0.8 & 0.8 &  \\
         optics efficiency $\epsilon_o$ & 0.3  & 0.35 & 0.7 & 0.7 & \\
         aperture $A$ & 30 & 7700 & 23000 & 39000 & cm$^2$ \\
         detector size $a$ & 0.15 & $2\times 0.3$ & $8\times 0.15$ & $8\times 0.15$ & cm$^2$ \\
         tracking fraction $\epsilon_t$ & 1 & 0.5 & 0.5 & 0.5 &  \\
         running time $T$ & 0.13  & 1.5 & 3 & 5 & year \\
         dark counting $\Phi_{\rm bkg}$ & $10^{-6}$ & $10^{-7}$ & $10^{-8}$ & $10^{-9}$  & $\mathrm{keV^{-1}\, cm^{-2} \, s^{-1}}$\\         
    \hline
    \end{tabular}
    \caption{The helioscope specifications we used for the analysis. We adapt the benchmarks from Refs.~\cite{Kotthaus:2005zg,Kuster:2007ue,CAST:2017uph,IAXO:2019mpb,IAXO:2020wwp}. In particular, the running time of CAST is taken to be the actual data-taking time, so $\epsilon_t$ is effectively unity there.}
    \label{tab:helioscope-specs}
\end{table}

We show in Fig.~\ref{fig:xi-ratio-bound} the improvement in terms of the enhancement factor $\xi(m_a)$ that AMR brings to these four experiments, compared to their sensitivity reach in the vacuum phase.
As a benchmark for the AMR enhanced search, we assume a helical magnetic profile that has a helical period of $2.4~\mathrm{m}$, which is motivated by the existing magnet technology of RHIC Snake magnet~\cite{zotero-21781,zotero-21778,zotero-21777,Anerella:2003se}.
The height of the peaks is $\sim 3$, with slight dependence on the experiments.
The location of the peaks is determined by $m_a \simeq \sqrt{2\omega \dot{\theta}}$, with typical photon energies $\omega\simeq \mathcal{O}(1)\,{\rm keV}$.
We also confirm the limiting behaviours expected from the conversion probability in Eq.~\eqref{eq:conversion-prob-varyingB}.
In the limit of $m_a^2/(2\omega) \gg |\dot{\theta}|$, the helical period becomes negligible, so the enhancement factor approaches 1.
In the opposite limit, the large $\dot{\theta}$ term appears in the denominator and suppresses the conversion probability, resulting in the enhancement factor vanishing.

\begin{figure}[t]
\centering
\begin{subfigure}{.49\textwidth}
\includegraphics[width=\linewidth]{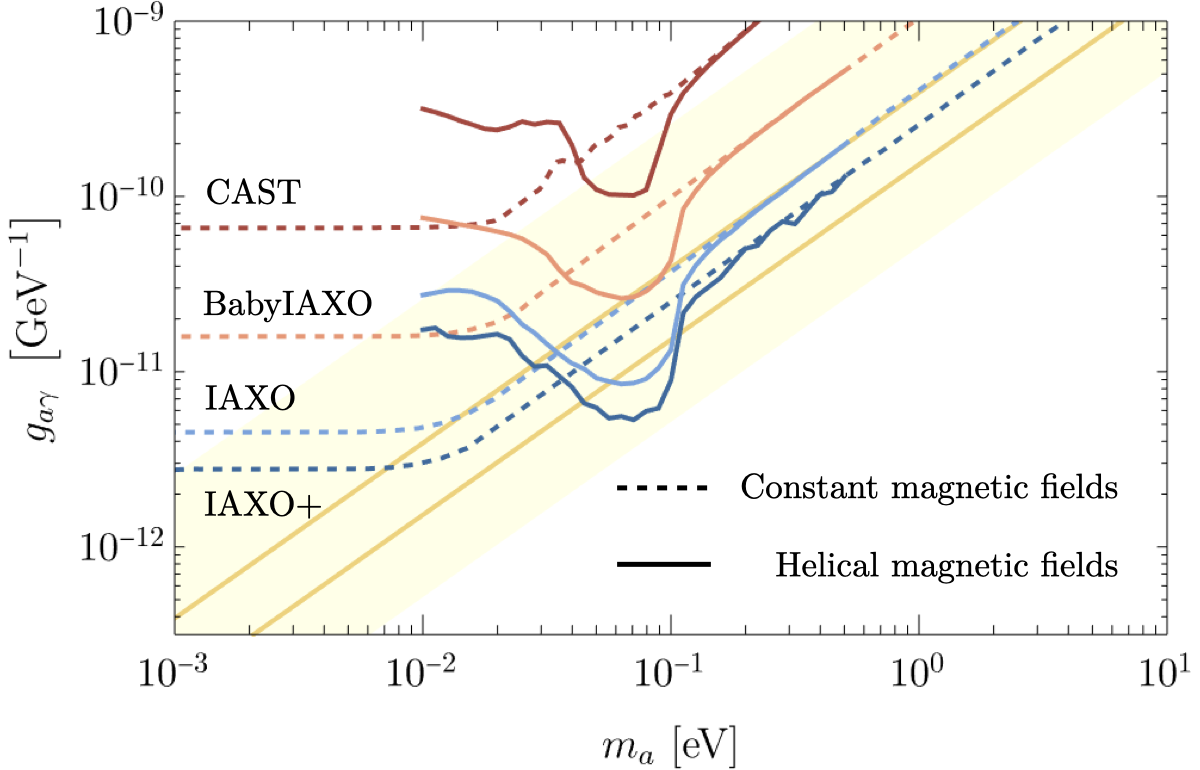}
\end{subfigure}
\begin{subfigure}{.49\textwidth}
\includegraphics[width=\linewidth]{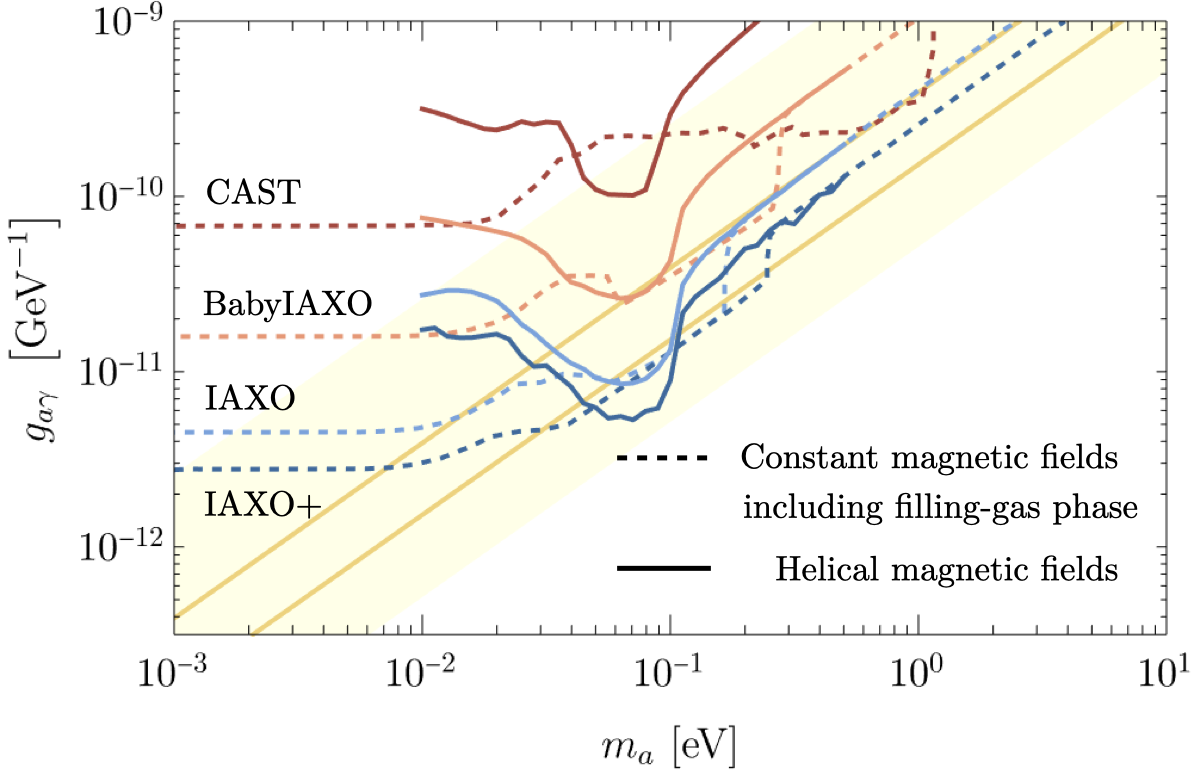}
\end{subfigure}
\caption{We show the improvement from AMR at CAST and projections for IAXO experiments.
We compute the experimental reach in $g_{a\gamma}$ based on the specifications in Table~\ref{tab:helioscope-specs}.
In this AMR benchmark, we assume the magnetic field rotates with a period of $2.4$ meters, a helical profile achieved by the RHIC Snake magnet, while having the same magnetic field strength with the constant magnetic field counterpart for each experiment (see Table~\ref{tab:helioscope-specs}).
The official bounds are presented in dotted lines without (left) and with (right) gas-induced plasma frequency~\cite{CAST:2017uph}.
Also shown is the QCD axion band with two central lines of KSVZ ($E/N=0$) and DFSZ ($E/N=8/3$) axion models.
The edges of the QCD axion band are set by $E/N=44/3$ (upper) and $E/N=5/3$ (lower).}
\label{fig:AMR-combined}
\end{figure}

In Fig.~\ref{fig:AMR-combined}, we show the experimental bounds in the ALP parameter space with (solid) and without (dotted) AMR.
The constraints without AMR, using constant magnetic fields, are taken from the literature.
The right panel additionally shows the constraint extension by including the filling-gas phases.
To better control over experimental details such as noise and efficiencies, instead of directly plotting our $g_{a\gamma}$ constraints, we multiply the enhancement factor $\xi(m_a)$ by the constraints from the literature, especially in Fig.~\ref{fig:AMR-combined}. {In particular, for CAST, we use the combined data from 2003-2011 and the analysis performed in 2017.}
We compare results from our simplified analysis with those from the literature as a sanity check (see Fig.~\ref{fig:analysis-comparison} in App.~\ref{sec:comparison-with-official}).
We successfully reproduce the official bounds to good precision. 

\subsection{Experimental Relevance}

\begin{figure}[h]
  \centering
  \includegraphics[width=.95\linewidth]{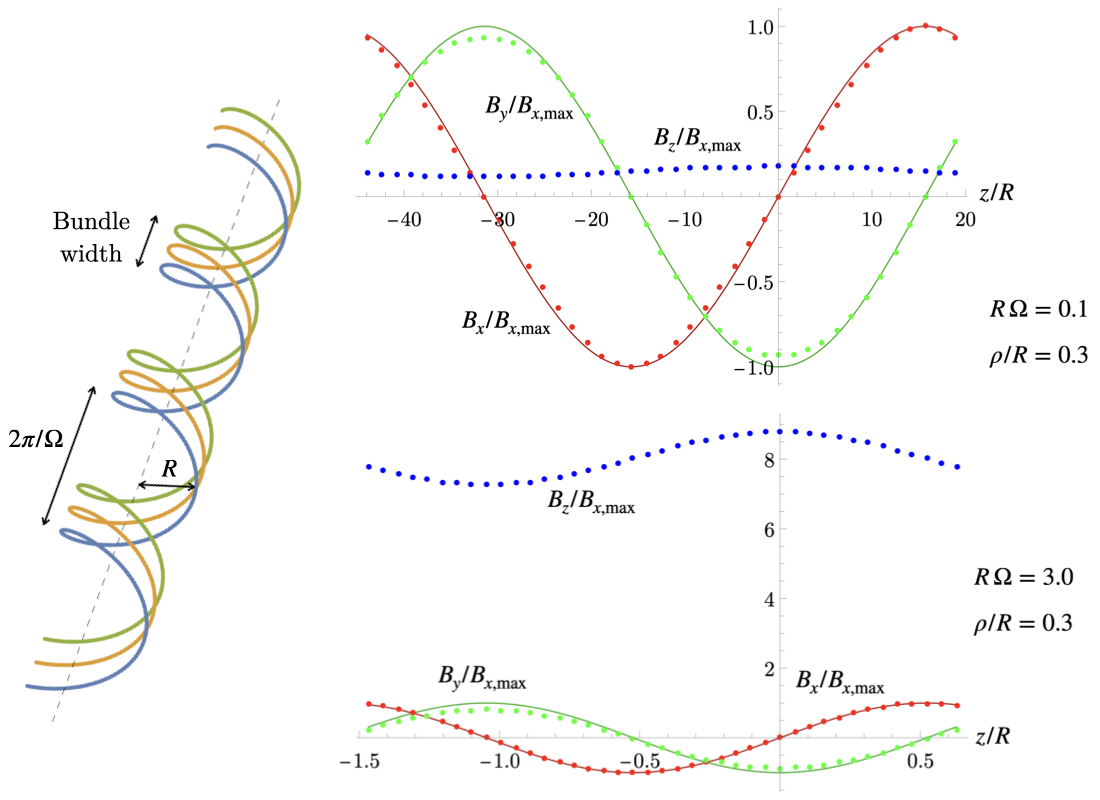}
  \caption{Schematic picture of helical currents (left), within which generate helical magnetic fields around the axis.
  The right panels represent the resulting magnetic field profiles for helical currents, depending on the parameters of helical currents.
  The results show that the $B_x$ (red) and $B_y$ (green) components reproduce the sine and cosine forms (solid lines) well.
  The $B_z$ component is colored blue.
  The relevant parameters are defined with respect to the radius $R$.
  (Right top) For $R\,\Omega=0.1$, the magnetic field is calculated at the radial coordinate $\rho = 0.3 R$. The number of current cables, $N_{\rm cab}$, is set to $5$, within a bundle width of approximately half the spatial period of the current, $0.5\times (2\pi/\Omega)\times(N_{\rm cab}-1)/N_{\rm cab}$.
  (Right bottom) Same as the top panel, but with $R\,\Omega = 3.0$.
  }
  \label{fig:bundle}
\end{figure}

Since we utilize spatially varying magnetic fields, it is essential to establish the feasibility of these configurations before advancing to the next section.
First of all, mature technologies for helical magnets, such as those used in the RHIC Snake magnet~\cite{zotero-21781,zotero-21778,zotero-21777,Anerella:2003se}, already exist.
Likewise, one way to realize a helical magnetic field profile is employing helical currents.
To demonstrate this idea, we calculate the magnetic field profiles generated by helical currents, as shown in Fig.~\ref{fig:bundle}, using the Biot-Savart law.
The position vector for a single helical current can be parametrized as $\mathbf{l'}=(R\cos\phi',R\sin\phi',\phi'/\Omega)$, where $\Omega$ is the spatial frequency of the helical currents, resulting in the magnetic fields with $\dot\theta=\Omega$.
We can factor out the radius $R$ and define all length scales relative to $R$.
Up to normalization, we find that the magnetic fields exhibit a helical profile around the central axis of the currents.
To broaden the region where the magnetic fields closely follow the helical profile, we introduce additional currents to form a helical bundle.
As shown in Fig.~\ref{fig:bundle}, at the radial coordinate $\rho=0.3 R$, we find the magnetic fields still follow the helical profile.
To satisfy Maxwell's equations, the $B_z=B_z(x,y,z)$ component appears, which does not impact our sensitivity.

\begin{figure}[t]
  \centering
  \includegraphics[width=0.6\linewidth]{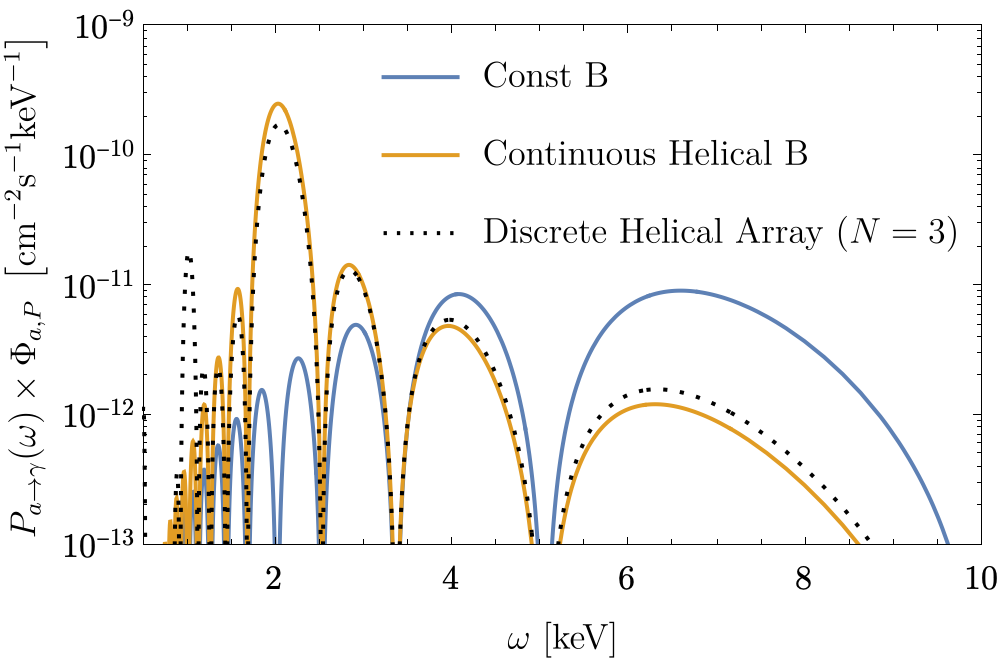}
  \caption{Similar to Fig.~\ref{fig:benchmark-flux}, but showing the case of a discrete helical array with $N=3$.
  For this configuration, the constant magnetic field is rotated by $2\pi/N$ at every $(2\,{\rm m})/N$ along the propagation direction to mimic the continuous helical profile.
  }
  \label{fig:Discrete}
\end{figure}

We also examine a discrete setup to relax the requirements.
We define a discrete helical array with an integer $N$ as a series of constant magnetic field units having rotated by a fixed angle of $2\pi/N$ relative to the previous one.
In this definition, alternating dipole magnets, referred to as wigglers or undulators, correspond to a specific case of $N=2$.
The photon spectrum converted from solar axions as Fig.~\ref{fig:benchmark-flux} is derived for the magnetic fields of a discrete helical array with $N=3$ in Fig.~\ref{fig:Discrete}.
The results indicate that the $N=3$ configuration already produces promising signals comparable to those of the continuous helical magnetic field configuration.
Even we observe that the $N=2$ case yields sufficiently strong signals since the magnetic fields contain a significant helical contribution within their Fourier modes.
To sum up, both continuous and discrete realizations of helical magnetic fields exist and provide robust benchmarks to adopt these setups for our purpose.

\section{Distinguish QCD Axion Models with AMR}
\label{sec:qcd-axion}

In the previous section, we demonstrated how the AMR method enhances the sensitivity of a helioscope, allowing it to explore a broader parameter space for the interaction between axion-like particles and photons.
We now focus on the QCD axion scenarios as the most motivated models for axions; see Refs.~\cite{GrillidiCortona:2015jxo,Agrawal:2017cmd,Hook:2018dlk,DiLuzio:2020wdo,Reece:2023czb} for some nice reviews of QCD axions' properties and beyond.

While QCD axions enjoy the same AMR enhancement, we stress that the solar axion flux differs in different QCD axion models. This is due to the relation between the axion-electron $g_{ae}$ coupling and axion-photon coupling $g_{a\gamma}$. Therefore, we study the AMR-enhanced experimental reach in both DFSZ and KSVZ models. 


\subsection{Brief Review of Axion EFT}
\label{sec:qcd-axion-review}

For phenomenological approaches, one needs to understand how QCD axions interact with the SM contents in low-energy scales.
In general, QCD axion models predict additional low-energy effective axion couplings to the SM particles apart from photons, such as the axion-electron and axion-nucleon couplings.
All these interactions contribute to the solar axion flux from the Sun, on top of the flux from the Primakoff process, the rate of which is derived in Eq.~\eqref{eq:primakoff}.
 
In the non-linear PQ symmetric basis, where the axion field shifts by a constant under the PQ transformation, the coupling of QCD axions to a fermion $f$ is expressed as a derivative term
\begin{align}
\mathcal L
& \supset
-\frac{g_{af}}{2}
{\partial_\mu a} \,
\bar f \gamma^\mu \gamma^5 f\,.
\label{eq:axion-fermion_coupling}
\end{align}
The dimensionful coefficient $g_{af}$ can be parametrized as $g_{af} \equiv c_{af}/f_a$, where $f_a$ denotes the decay constant normalized in terms of the gluon anomaly, and $c_{af}$ is associated with the PQ charge of the fermion.
Note that the field basis for describing interactions of QCD axions can be redefined by an axionic $U(1)$ rotation of fields, which relies on the divergence of currents (i.e., current conservation).
Since the axial vector current contains a non-vanishing divergence due to its mass\footnote{The divergence of an axial vector current may have an anomalous term for photons. However, as shown below, the effective axion-photon coupling
is independent of the field redefinition as it should be.}, the QCD axion couplings in Eq.~\eqref{eq:axion-fermion_coupling} correspond to the non-derivative couplings as $g_f m_f a\bar{f}i\gamma^5f$ in the linear order.
Hence, one can easily match dimensionless axion non-derivative couplings used in the literature to our notation with dimensionful parameters in Eq.~\eqref{eq:axion-fermion_coupling}.
Similarly, vector current couplings are conserved in the low-energy regime, so they are rotated away and do not affect low-energy processes.

In a field-theoretical model, the low-energy effective QCD axion coupling to photons receives contributions from three main components.
Firstly, there can be UV contributions arising from the topological term of photons, $(\alpha/8\pi)F\tilde F$, where its coefficient in terms of $a/f_a$ is strictly quantized (i.e., the so-called Wilson coefficient).
Additionally, there are two corrections in the IR domain from mixing with mesons and that from triangle loops involving fermions charged under both the electromagnetic and PQ symmetries.
Overall, we can parametrize the axion-photon effective coupling to be 
\begin{align}
g_{a\gamma, \rm eff}
& = 
g_{a\gamma, \rm UV} 
+ g_{a\gamma, \chi{\rm PT}}
+ g_{a\gamma, \rm \ell}\,.
\end{align}

The model-dependent UV contribution, $g_{a\gamma, \rm UV}$, arises from all the PQ-charged fermions with a non-vanishing electric charge.
Given that the gluon anomaly with respect to QCD axions is normalized as $(a/f_a)(\alpha_s/8\pi)G\tilde{G}$, $g_{a\gamma, \rm UV}$ reads
\begin{align}
\label{eq:gagammaUV}
    g_{a\gamma, \rm UV} =\frac{\alpha}{2\pi f_a}
\left (
\frac{E}{N} \right )\,,
\end{align}
where $N$ denotes the domain wall number, $E=2\,{\rm tr}[c_{af} q_f^2]$, $c_{af}$ and $q_f$ are the QCD axion coupling coefficient to the PQ current and the electric charge of a fermion $f$, respectively, and the trace is taken over all the fermions and its internal degrees of freedom.

The anomalous strong interaction of QCD axions, which is essential for addressing the strong CP problem, induces the mixing with mesons at scales below the confinement scale.
The diagonalization of these mixings leads to a model-independent contribution to the axion-photon coupling $g_{a\gamma}$.
Based on the chiral perturbation theory, which assesses mixings between QCD axions and pseudoscalar mesons, we find 
\begin{align}
\label{eq:gagammachiPT}
    g_{a\gamma, \chi{\rm PT}}
& \approx
-\frac{\alpha}{2\pi f_a} 
\left( \frac{2}{3}
\frac{4 m_d + m_u}{m_u + m_d}
\right ) \,. 
\end{align}
This result is dominated by the mass mixing with neutral pions, while contributions from kinetic mixings become negligible due to the small QCD axion mass compared to mesons.


There are additional corrections to the axion-photon coupling in the IR by loops involving charged lepton.
In the presence of the QCD axion interaction to a charged lepton $l$ as in Eq.~\eqref{eq:axion-fermion_coupling}, $(-g_{al}\partial_\mu a /2)\bar{l}\gamma^\mu\gamma^5 l$, the 1PI amplitude of QCD axions and photons at the characteristic energy scale of the process denoted by $\mu$ reads
\begin{align}
    g_{a\gamma, \ell}
    & = \sum_l \frac{\alpha}{\pi}g_{al} \times\left[-1 + \frac{1}{\tau_l}
    \begin{cases}
    -\frac{1}{4}\left(\log\frac{1+\sqrt{1-\tau_l^{-1}}}{1-\sqrt{1-\tau_l^{-1}}}-i\pi\right) & \tau_l > 1  \\
   {\rm arcsin}^2\sqrt{\tau_l}
    & \tau_l \leq 1
    \end{cases}\right] \,
\end{align}
with $\tau_l = \mu^2/4m_l^2$.
Since the energy scale of interest is in the keV range, which is much smaller than the mass of the lightest lepton (i.e., electron), 1PI corrections are negligible in our analysis.

By setting the model dependent parameters, such as $c_{af}$ and $E/N$, to be in the natural order of $\mathcal O(1)$, we see that $g_{a\gamma} \sim 10^{-3} /f_a$ while $g_{af} \sim 1/f_a$.
Consequently, the solar axion flux due to $g_{ae}$ can be orders of magnitude larger than that from $g_{a\gamma}$~\cite{Barth:2013sma,Redondo:2013wwa,Jaeckel:2018mbn}.
It is worth noting that while the axion-photon coupling does not rely on $g_{af}$, the flux's dependence on $g_{af}$ provides axion helioscopes with an additional means to distinguish different QCD axion models~\cite{Jaeckel:2018mbn}.


There are two types of QCD axion models: KSVZ~\cite{Kim:1979if,Shifman:1979if} and DFSZ~\cite{Dine:1981rt,Zhitnitsky:1980tq}.
In KSVZ models, the gluon anomaly is realized by introducing heavy PQ colored fermions.
For simplicity, we consider heavy fermions that are singlet under the electroweak symmetry, resulting in $g_{a\gamma} \simeq -1.92 \alpha/2\pi f_a$.
Since there is no tree-level coupling of QCD axions to electrons, the solar axion flux and its detection rate in helioscopes are governed by the axion-photon coupling.
On the other hand, DFSZ models extend the scalar sector of the SM, with SM fields charged under the PQ symmetry.
Thus, the solar axion flux is significantly influenced by the tree-level axion-electron coupling.
Building upon (the specific type of) two Higgs doublet model as the simplest extension of SM, the axion-electron coupling is given by $g_{ae} = \sin^2\beta/3f_a$, where $\beta$ parametrizes the ratio between the vacuum expectation values of Higgs doublets defined as $\tan\beta = v_u/v_d$.
This $\beta$ dependence arises from the $Z$-boson threshold effect.
We adopt a large $\tan\beta$ value, which is typically relevant for phenomenological contexts, resulting in $g_{ae} \simeq 1/3f_a$.
The additional model dependence is embedded in the $E/N$ value for $g_{a\gamma}$, the boundary values of which are $5/3$ and $44/3$.
As we will discuss in the next subsection, although there are tree-level couplings to nucleons in both models, their contributions to the solar axion flux in the keV range are sub-dominant due to the heavy nucleon mass.

\subsection{Solar Flux of QCD Axions}
\label{sec:flux-qcd}

Aside from the Primakoff process, 
we now examine the contributions to the solar axion fluxes in a keV range by axion-fermion couplings. 

~\\\textbf{Electrons}~~ We have two main scattering processes for the coupling to electrons: Compton-like and bremsstrahlung (free-free).
The respective spectral fittings can be expressed as follows~\cite{CAST:2007jps,Barth:2013sma}
\begin{align}
\Phi_{a, C}
& = 
3.5\times 10^6~
\left ( \frac{g_{ae}}{10^{-10}\,\mathrm{GeV}^{-1}} \right )^2
\frac{\tilde \omega^{2.987}}{e^{0.776\,\tilde \omega}}
~~
[\mathrm{cm}^{-2}\, {\rm s}^{-1} \,
\mathrm{keV}^{-1}],
\label{eq:PhiaCompton}
\\
\Phi_{a,B}
& = 
6.9\times 10^8 \,
\left ( \frac{g_{ae}}{10^{-10}\,\mathrm{GeV}^{-1}} \right )^2
\frac{\tilde \omega}{1+0.667 \tilde \omega^{1.278}}
\mathrm e^{-0.77 \tilde \omega}
~~
[\mathrm{cm}^{-2}\, {\rm s}^{-1} \,
\mathrm{keV}^{-1}].
\label{eq:PhiaBrem}
\end{align}
Additional processes related to electronic atomic transitions, such as atomic deexcitation (bound-bound) and recombination (bound-free), also contribute to the total flux.
Their contributions are smaller than those of bremsstrahlung by an $\mathcal O(1)$ factor; see Ref.~\cite{Redondo:2013wwa} for details.
To stay conservative, our analysis only include the solar axion flux generated by the Compton-like and bremsstrahlung processes.
Including atomic transition processes would enhance our results slightly by an $\mathcal{O}(1)$ factor.

~\\\textbf{Nucleons}~~The solar axion flux also includes contributions from the axion-nucleon couplings, which are present at the tree-level order for QCD axion models.

\textit{$\gamma  N \rightarrow N a$}~~~~ In the non-relativistic limit, the rate of Compton-like scatterings is inversely proportional to the mass square of the target particle.
Thus, Compton-like scatterings with nuclei give a negligible rate compared to scatterings with electrons.

There are two bremsstrahlung processes involving nucleons, electron-nucleon and nucleon-nucleon scatterings.
In these processes, the QCD axion is now attached to a nucleon line in the diagrams.
For both non-relativistic electrons and nucleons, their typical momentum is much larger than the temperature, allowing the axion momentum to be effectively ignored in momentum conservation of kinematics.
Furthermore, since nucleons are much heavier than electrons, the momentum transfer in electron-nucleon bremsstrahlung is dominated by electrons, accounting for nucleons as almost relatively fixed objects during an interaction.

\textit{$e  N \rightarrow eN^* \rightarrow eNa$}~~~~In comparison with electron-nucleon bremsstrahlung via the axion-electron coupling (\textit{i.e.} $e  N \rightarrow e^*N \rightarrow eNa$), the phase space integrals are the same, but the scattering amplitude squared is suppressed by approximately $(c_N m_N/c_e m_e)^2(m_e^4/m_N^4)$.
The first and second terms of this suppression arise from the coupling strength and the mass dependence of the intermediate particle (i.e., the fermion attached to the axion) and the target, respectively.
This indicates that the rate of electron-nucleon bremsstrahlung via the axion-nucleon coupling is also significantly reduced by the nucleon mass squared.

\textit{$N  N \rightarrow NN^* \rightarrow NNa$}~~~~In nucleon-nucleon bremsstrahlung, the momentum transfer is of an order of the nucleon momentum, resulting in a different mass scaling behavior.
According to Refs.~\cite{Raffelt:1985nk,Redondo:2013wwa}, its rate is proportional to $(c_N m_N/f_a)^2 v_N^5 m_N^{-1}$ with the nucleon velocity $v_N \sim \sqrt{T/m_N}$.
Although nucleon-nucleon bremsstrahlung is less suppressed (scaling as $ m_N^{-3/2}$) compared to other processes above, we find that the contribution of the axion-nucleon couplings to the axion flux is still much less than that from the axion-electron coupling and even that from the axion-photon coupling when the dimensionless model-dependent parameters are $\mathcal{O}(1)$.

Lastly, let us look at the nuclear decay and transition lines, which is the nuclear counterpart of the free-bound and bound-bound atomic transition processes.
Historically, there have been active searches for axion fluxes from both processes.
In the nuclear decay process, the axion flux is expected to be at the MeV range~\cite{CAST:2009klq}. The $^7\mathrm{Li}^* \rightarrow ^7\mathrm{Li} + a$ decay gives an axion flux at 0.478~MeV around $10^{-11} (f_a/10^{-10}\,\mathrm{GeV}^{-1})^2\,\mathrm{cm}^{-2} \mathrm{s}^{-1}\,\mathrm{keV}^{-1}$ with a width about 0.2~keV at the solar core. The $p+d \rightarrow ^3\mathrm{He} + a$ process generates an axion flux at 5.5~MeV around $10^{-9} (f_a/10^{-10}\,\mathrm{GeV}^{-1})^2\,\mathrm{cm}^{-2} \mathrm{s}^{-1}\,\mathrm{keV}^{-1}$. Therefore, we can neglect these contributions. 
At lower energy, the thermally excited of stable isotopes generate axion flux at the keV range. The two popular candidates include $^{56}\mathrm{Fe}$~\cite{Moriyama:1995bz,Krcmar:1998xn,CAST:2009jdc,zotero-22774,XENON:2020rca,DiLuzio:2021qct} and $^{83} \mathrm{Kr}$~\cite{Gavrilyuk:2014mch} both have a spectral flux much smaller than that from the electron bremsstrahlung. For example, $^{56}\mathrm{Fe}$ has a flux at 14.4~keV with a width $\sim 5\,\mathrm{eV}$, $\Phi \approx 8.9\times 10^{5}\, (f_a/10^{-10}\,\mathrm{GeV}^{-1})^2\,\mathrm{cm}^{-2} \mathrm{s}^{-1}\,\mathrm{keV}^{-1}$. 
As a result, in what follows, we neglect the flux component due to the axion-nucleon coupling.




\begin{figure}[th]
\centering
\begin{subfigure}{.49\textwidth}
\includegraphics[width=\linewidth]{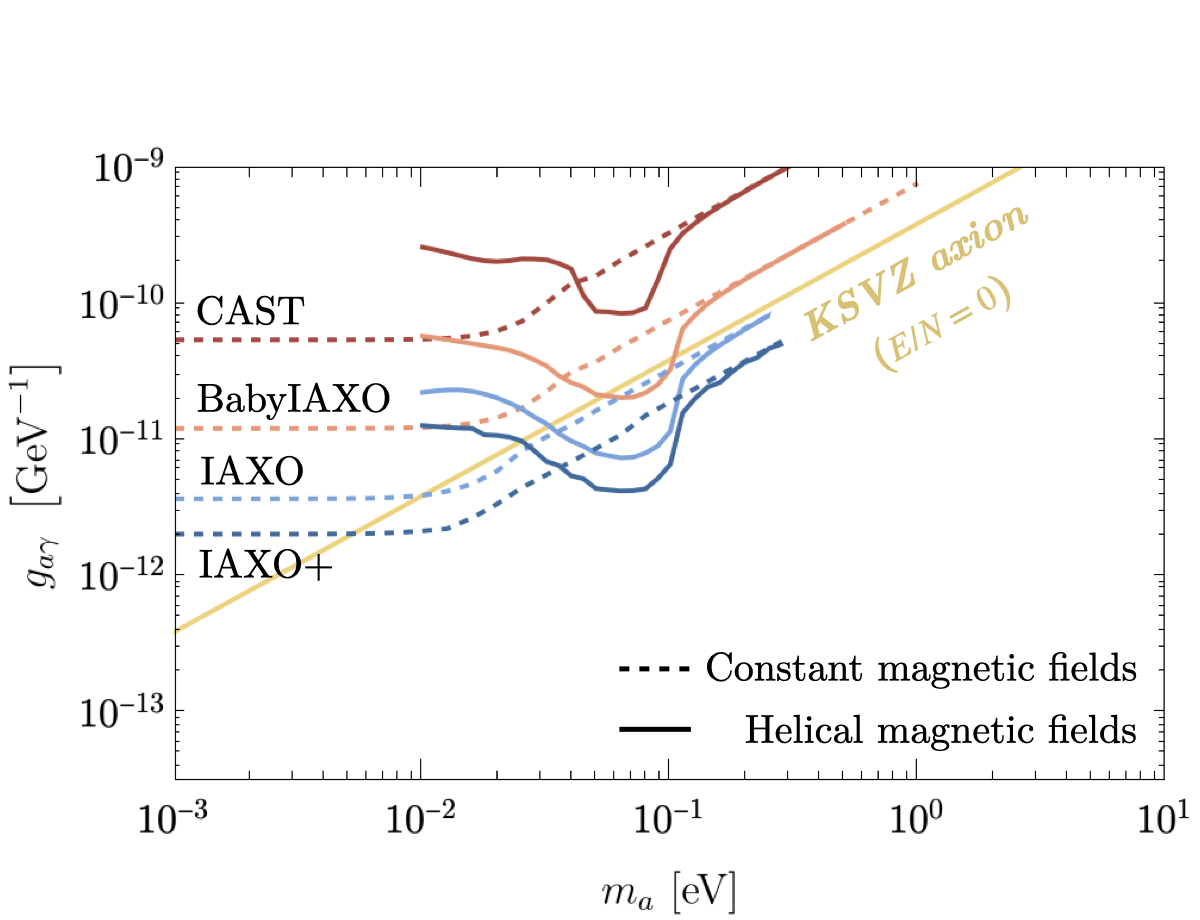}
\end{subfigure}
\begin{subfigure}{.49\textwidth}
\includegraphics[width=\linewidth]{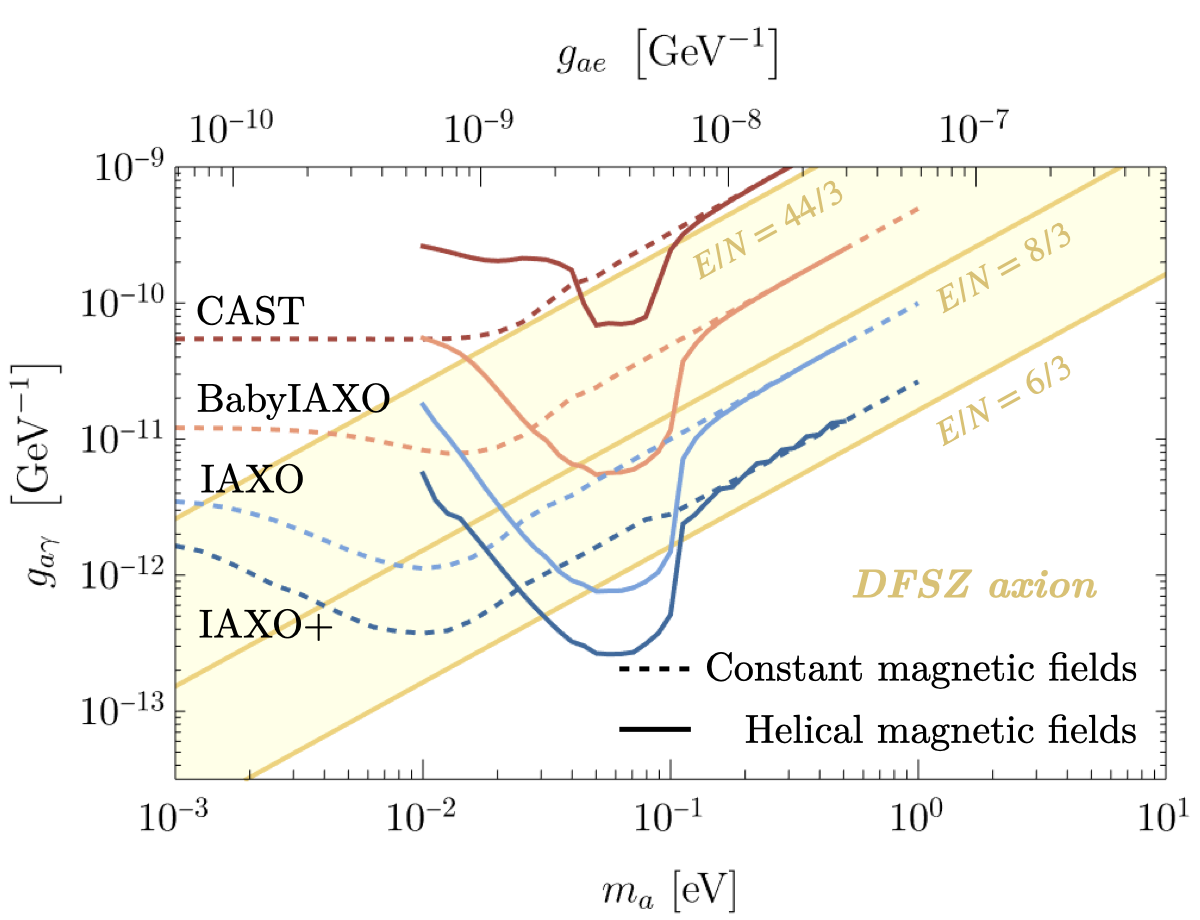}
\end{subfigure}
\caption{
We perform the extended maximum likelihood analysis on QCD axion models with (solid) and without (dotted) AMR. We adopt the experimental specifications same as Fig.~\ref{fig:AMR-combined}.
\textbf{Left}: in the KSVZ axion case, we show the improvement AMR brings to CAST (red), which approaches the KSVZ line, and to BabyIAXO (orange), IAXO(light blue), and IAXO+ (darker blue), which surpass the KSVZ line with significance.
\textbf{Right}: in the DSFZ case, $g_{ae}$ is a function of the axion mass for a given $\tan \beta=v_u/v_d$. We consider $\tan\beta\gtrsim \mathcal{O}(1)$ as a benchmark, resulting in $g_{ae} \simeq 1/(3f_a)$. $g_{a\gamma}$ varies due to the model-dependent choice of anomaly coefficients $E/N$, hence the yellow band of target theories.
}
\label{fig:QCDaxions}
\end{figure}

\subsection{Helioscope Sensitivities on QCD Axions with AMR}
\label{sec:sensitivity-qcd}

In Fig.~\ref{fig:QCDaxions}, we present the helioscope sensitivities for QCD axions with and without AMR, utilizing our simplified analysis as in Sec.~\ref{sec:reso-with-extend}.
We examine two benchmark models for QCD axions: the KSVZ and DFSZ models.
As discussed in Sec.~\ref{sec:qcd-axion-review}, the QCD axion mass $m_a$ is the defining parameter for axion couplings, such as to photons and electrons, with an additional degree of freedom $E/N$.
We thus include solar axions produced not only from the Primakoff process via the axion-photon coupling $g_{a\gamma}$ but also through the Compton-like process and bremsstrahlung via the axion-electron coupling $g_{ae}$, as given in Eqs.~\eqref{eq:PhiaCompton} and \eqref{eq:PhiaBrem}.

The left panel of Fig.~\ref{fig:QCDaxions} illustrates the helioscope sensitivities for the KSVZ model with $E/N = 0$, corresponding to the yellow line.
Since the axion-electron coupling is highly suppressed in the KSVZ model due to the absence of tree-level UV contributions, solar axions from electron-induced processes are negligible compared to those from the Primakoff process, and only the axion-photon coupling is relevant across the parameter space.
Therefore, the KSVZ model provides almost the same constraints as those shown in Fig.~\ref{fig:AMR-combined}, where only the axion-photon coupling is considered, up to only negligible differences due to the simplified assumption in Eq.~\eqref{eq:Nexpassump} (see Fig.~\ref{fig:analysis-comparison}).\footnote{
Since the constraints on $g_{a\gamma}$ in the literature do not account for the axion-electron coupling contribution, we tolerate the negligible differences rather than multiplying the enhancement factor $\xi(m_a)$ by the reference values from the literature, as done for AMR in Sec.~\ref{sec:amr-at-helioscopes}.
}

The helioscope sensitivities for the DFSZ models are presented in the right panel of Fig.~\ref{fig:QCDaxions}.
As summarized in Secs.~\ref{sec:qcd-axion-review} and \ref{sec:flux-qcd}, the axion-electron coupling in the DFSZ model significantly boosts the solar axion flux compared to the flux generated solely by the axion-photon coupling.
For our benchmark, we consider $\tan\beta \gtrsim \mathcal{O}(1)$, which results in $g_{ae}\simeq 1/(3f_a)$.
The corresponding $g_{ae}$ values are shown on the top axis, which aligns with the respective $m_a$ in the context of QCD axions.
We set $\dot{\theta} = 2\pi/(2.4\,{\rm m})$ and use the other specifications listed in Table~\ref{tab:helioscope-specs}.
The AMR manifestly improves the sensitivity and enables each experiment to reach $E/N$ lines that were previously unattainable.
From a different perspective, the AMR substantially increases the signal-to-noise ratio or reduces the required running time of experiments around specific axion masses, which depend on $\dot{\theta}$; for instance, $m_a \simeq 10^{-1}\,{\rm eV}$ for $\dot{\theta} = 2\pi/(2.4\,{\rm m})$.
To highlight our improvement, we scan $E/N$ down to $E/N=6/3$, where a subtle cancellation between $g_{a\gamma, \rm UV}$ and $g_{a\gamma, \chi{\rm PT}}$ in Eqs.~\eqref{eq:gagammaUV} and \eqref{eq:gagammachiPT} results in an order-of-magnitude suppression of $g_{a\gamma, \rm eff}$.\footnote{Note that the $E/N=5/3$ line lies between $E/N=8/3$ and $E/N=6/3$ lines because we are interested in the absolute value of $g_{a\gamma, \rm eff}$.}
Our AMR approach successfully probes this photo-phobic line, even with the IAXO setup.
One additional remark on the case without AMR (dotted line) is a dip, instead of a plateau, on the left side of the contour. 
This dip is due to the increase in the axion-electron coupling as the axion mass increases.

\section{Conclusion and Outlook}
\label{sec:conclusion-outlook}

Axions and axion-like particles are hypothetical particles that appear ubiquitous in various UV-completed models and, more importantly, involve compelling motivations from a phenomenological perspective.
Detecting these particles and uncovering their properties remain one of the central goals in the particle physics community.
However, due to their pseudo-scalar natures, axions are expected to be naturally light with very feeble interactions with the SM contents, making their detection highly challenging.
While many ongoing and planned experiments are focused on enhancing detector precision and sensitivity to explore a broader parameter space, the development of innovative experimental methodologies could provide a breakthrough needed to overcome such challenges.

In the framework of the effective theory, the characteristic low-energy effective axion coupling is its anomalous interaction with photons, $(g_{a\gamma}/4)aF_{\mu\nu}\tilde{F}^{\mu\nu}$.
This coupling allows for a fascinating phenomenon: in the presence of an electromagnetic background, axions can mix with photons which leads to their mutual conversion.
Many axion detection experiments leverage this axion-photon conversion mechanism.
For example, in the Light-Shining-Through-Walls (LSTW) experiments, axions produced by incident photons can pass through solid barriers, and the subsequent regeneration of photons via the conversion of such axions serves as detectable signals.
In conventional setups with a constant magnetic field background, sensitivities are lost for higher axion masses, where the axion-photon oscillation length becomes much shorter than the scale of the magnetic field domain.
One promising approach to overcome this limitation is the use of a spatially varying magnetic field.
When the frequency of this variation matches the axion-photon oscillation frequency, sensitivity for detecting axions of the corresponding mass can be significantly enhanced.
This methodology is the so-called `Axion Magnetic Resonance'~\cite{Seong:2023ran}.

We extended the application of the AMR idea to helioscopes, which detect axions sourced by the Sun, and assessed the resulting improvement in sensitivity.
Similarly, conventional helioscopes rely on the constant magnetic field background for photon regeneration in detectors. which presents challenges in probing axions with masses above $10^{-2}\,{\rm eV}$.
AMR can address this difficulty by enhancing sensitivities at the heavy mass regime, particularly around $0.1\,{\rm eV}$ for the existing helical profile of RHIC Snake magnet as shown in Fig.~\ref{fig:AMR-combined}.
Furthermore, since the solar axion flux arising from the Primakoff process spans a broad spectrum between $1\text{-}10\,{\rm keV}$, AMR allows us to explore a wide range of axion parameter space with significantly improved sensitivity, even using a single helical magnetic field configuration.

Effective low-energy axion couplings to particles other than photons, such as electrons and nucleons, are generically present in usual field-theoretic axion models, where axions originate from the phase of PQ-charged scalar fields.
While the interaction of axions to photons is anomalous and radiatively induced at one-loop order, the axion couplings to electrons and nucleons may appear at the tree-level order, as associated with their effective PQ-charge, which can significantly contribute to the solar axion flux.
As specific but well-motivated examples, we examined QCD axion scenarios.
Particularly in DFSZ-type QCD axion models, the axion coupling to electrons is given in natural order in terms of $f_a^{-1}$ that dominates the solar axion flux; the contributions from similar tree-level couplings to nucleons are much suppressed by the relatively large nucleon mass as discussed in Sec.~\ref{sec:flux-qcd}.
We then investigated the sensitivities of helioscopes to search for QCD axions with AMR.
In the assumption of the correlated axion-electron coupling strength with the axion mass as in DFSZ axion scenarios, we realized that future helioscopes with a single helical magnetic setup could cover the entire DFSZ axion parameter space for masses in the range of $10^{-2}\text{-}10^{-1}\,{\rm eV}$, as shown in the right panel of Fig.~\ref{fig:QCDaxions}.

We highlight that the concept of AMR has broader applications beyond laboratory searches, extending to cosmological and astrophysical implications of axions.
In reality, astrophysical and cosmological magnetic field backgrounds are not constant, but varying spatially (and also temporally) in space.
Such a variability introduces a space- and time-dependent Hamiltonian that governs axion-photon oscillations along with the line of propagation, then this potentially leads to unique axion-photon oscillation patterns that differ from those predicted by models assuming a constant background profile.
Exploring signals associated with these varying backgrounds could yield intriguing new insights into probing axions.


\acknowledgments
HS is supported by the Deutsche Forschungsgemeinschaft under
Germany Excellence Strategy — EXC 2121 “Quantum
Universe” — 390833306. 
HS appreciates the hospitality of LAPTh during the visit.
 This work was partly supported by the U.S. Department of Energy through the Los Alamos National Laboratory. Los Alamos National Laboratory is operated by Triad National Security, LLC, for the National Nuclear Security Administration of U.S. Department of Energy (Contract No. 89233218CNA000001). Research presented in this article was supported by the Laboratory Directed Research and Development (LDRD) program of Los Alamos National Laboratory under projects 20220135DR and 20230047DR.
CS acknowledges the receipt of the grant from the Abdus Salam International Centre for Theoretical Physics (ICTP), Trieste, Italy, where part of this work was finished. 
CS thanks the hospitality and support of IBS-CTPU and support of CKC for the CTPU-CKC Joint Focus Program: Let there be light (particles) Workshop, where part of this work was performed. 
SY was supported by IBS under the project code, IBS-R018-D1

\appendix




\section{Extended Maximum Likelihood 
}
\label{sec:fit}

In this section, we adapt the procedure given in \cite{Barlow:1990vca} to the analysis of the CAST data. We briefly review the method in this section. 

Let us denote $\mathcal{F}_{\rm th}$ the theoretical expectation of the particle spectral flux density in the unit of $\rm cm^{-2}\cdot{\rm sec}^{-1}\cdot{\rm keV}^{-1}$. It includes both the backgrounds and axion-induced signals. 

Now imagine that an experiment has observed an ensemble of events labeled as $\mathcal X = \{1, 2, ..., N_{\rm tot}\}$. Each index corresponds to a unique event. Let us imagine binning the data set such that within each bin there is at most one event. Within each bin the number counting follows the Poisson distribution $P(\lambda, n)$, where $\lambda$ is the \textit{expected} number of events (\textit{i.e.} mean) and $n$ the actual number of events. Since the measured number of events is always zero or unity due to the aforementioned binning method, the likelihood for each bin is given by 
\begin{align}
\mathcal L_{j}^{0}  & = P(\lambda; 0) = \mathrm e^{-\lambda} ,
\qquad
\mathcal L_{k}^{1}    
 = P(\lambda; 1)
= \lambda \; \mathrm e^{-\lambda} ,
\end{align}
where
\begin{align}
\lambda
& = A \Delta t\, dE
\mathcal{F}_{\rm th}(E).
\end{align}
In the above, $A$ is the effective area of the detector, $\Delta t$ exposure time, $dE$ the width of the energy bin. 
The likelihood of finding such a realization $\mathcal X$ is given by
\begin{align}
    \mathcal L 
    & = \left(\prod_j \mathcal L_j^0\right)
    \left(\prod_k^{N_{\rm tot}} \mathcal L_k^1\right)
    \cr
& = 
\left(
\prod_{k=1}^{N_{\rm tot}}
A \Delta t\, dE
\mathcal{F}_{\rm th}(E_k)
\right)
\left(
\prod_{\rm total\,\,range}
e^{-A \Delta t\, dE
\mathcal{F}_{\rm th}(E)
}
\right).
\end{align}
The advantage of choosing such a likelihood is that, so far, it does not depend on any fixed binning method. 
The log-likelihood function is given by
\begin{align}
\ln \mathcal{L}
&=
\sum_{k=1}^{N_{\rm tot}}
\ln\left[
A \Delta t\, dE
\mathcal{F}_{\rm th}(E_k)
\right]
-
\mathcal{N}_{\rm th}
,
\end{align}
In practice, one can still coarse-grain the data set $\mathcal X$ such that $N_{\rm tot}
=
\sum_{i=1}^{n_{\rm bin}}n_i^{\rm count}
.
$
This leads to a log-likelihood that is easier more computationally efficient
\begin{align}
\ln \mathcal L
&\simeq
\sum_{i=1}^{n_{\rm bin}}
n_i^{\rm count}
\ln\left[
A \Delta t\, dE
\bar{\mathcal{F}}_{\rm th}(E_i)
\right]
-
\mathcal{N}_{\rm th},
\\
&=
A\Delta t\, \Delta E_{\rm bin}\sum_{i=1}^{n_{\rm bin}}
\mathcal{F}_i^{\rm count}
\ln\left[
A \Delta t\, \Delta E_{\rm bin}
\bar{\mathcal{F}}_{\rm th}(E_i)
\right]
-
\mathcal{N}_{\rm th}
+\ln\left[dE/\Delta E_{\rm bin}
\right]^{N_{\rm tot}}
,
\end{align}
%
In the first equality, we approximate that $\mathcal{F}_{\rm th}(E_k)\simeq \bar{\mathcal{F}}_{\rm th}(E_i)$.
Given the data, $N_{\rm tot}$ is constant with respect to the parameters of the model, as well as $A$, $\Delta t$, $\Delta E_{\rm bin}$, and $dE$.
The last constant term can be dropped even for analyzing the errors and finding the confidence level of the parameters.

\section{Comparing with Official Constraints}
\label{sec:comparison-with-official}
In this section, we reproduce the experimental constraints without the use of AMR. We use the simplified analysis as outlined in Sec.~\ref{sec:amr-at-helioscopes} assuming a constant magnet. 

\begin{figure}[th]
  \centering
  \includegraphics[width=.7\textwidth]{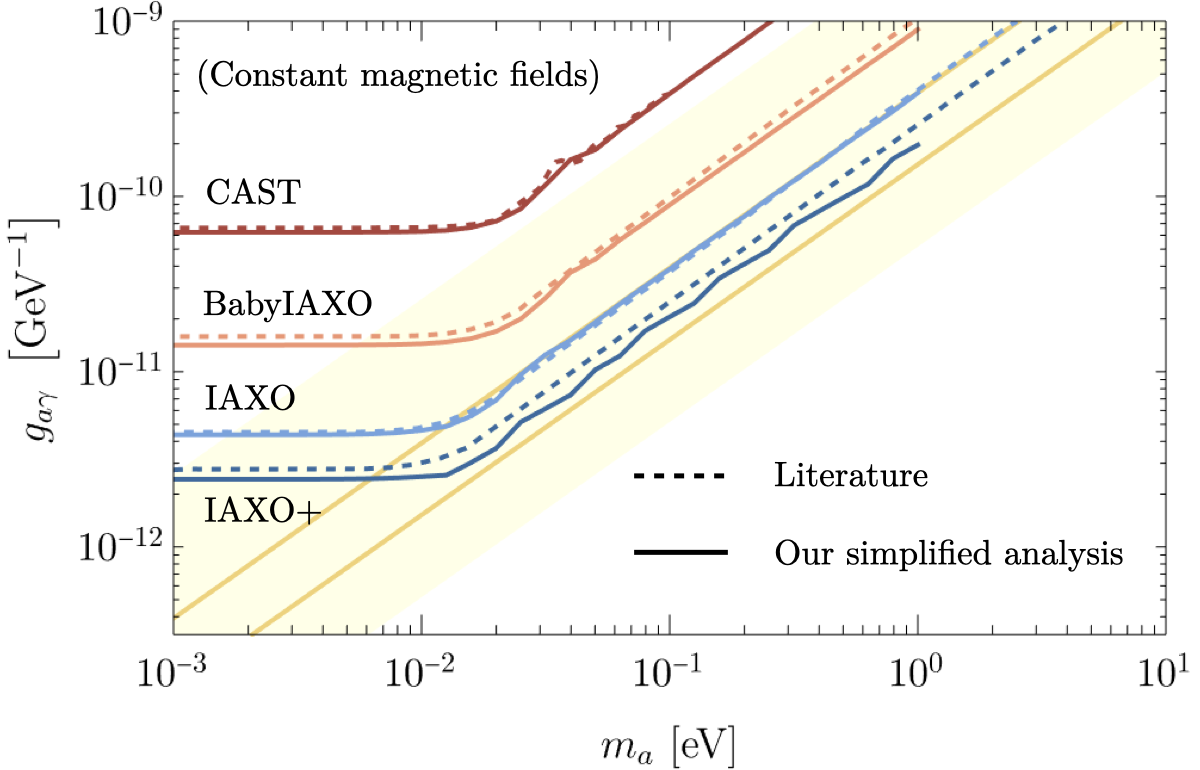}
  \caption{  We reproduce the official contours by applying the extended maximum likelihood method to the specifications outlined in Table~\ref{tab:helioscope-specs} with constant magnets. The solid curves are the results we reproduce, and the dotted curves are taken from Refs.~\cite{Kotthaus:2005zg,Kuster:2007ue,CAST:2017uph,IAXO:2019mpb,IAXO:2020wwp}. 
  \label{fig:analysis-comparison}
  }
\end{figure}

We then compare in Fig.~\ref{fig:analysis-comparison} our reproduced results with the officially released constraints and projections~\cite{Kotthaus:2005zg,Kuster:2007ue,CAST:2017uph,IAXO:2019mpb,IAXO:2020wwp}. We observe a negligible difference, which is likely due to the simplified background and detector modeling we assume. {For our purpose of self-consistency check, here we only use the official result of the analysis performed in 2017 for CAST.}

However, we stress that even such a small mismatch does not propagate into our final results. This is because the small difference only shifts the overall magnitude instead of the shape of the contours. As a result, it will not affect the AMR's enhancement factor $\xi(m_a)$ as defined in Eq.~\eqref{eq:enhancement-factor}. 
When we apply the enhancement factor to the official results to produce the results in Figs.~\ref{fig:AMR-combined} and \ref{fig:QCDaxions}, the overall shift is effectively factorized out.

\bibliography{bib}
\bibliographystyle{utphys}
\end{document}